\documentstyle[12pt]{article}

\begin{document} 
\begin{center}  {\Large {\bf Hadron Production in the
Trans-fragmentation Region in Heavy-Ion Collisions}}
\vskip .75cm
 {\bf Rudolph C. Hwa$^1$ and C.\ B.\ Yang$^{1,2}$}
\vskip.5cm
{$^1$Institute of Theoretical Science and Department of
Physics\\ University of Oregon, Eugene, OR 97403-5203, USA\\
\bigskip
$^2$Institute of Particle Physics, Hua-Zhong Normal
University, Wuhan 430079, P.\ R.\ China}
\end{center}

\vskip.5cm

\begin{abstract} 
We study the production of hadrons in Au+Au collisions in the region $0.6<x_F<1.2$, which we refer to as the trans-fragmentation region (TFR), since it corresponds roughly to $\eta'>0$, where $\eta'=\eta-y_{\rm beam}$, depending on $p_T$. We show how hadrons can be produced in that region when the hadronization process is parton recombination. The inclusive $x$-distributions for proton and pion production are calculated with momentum degradation taken into account.  The results show that the  proton yield is significantly higher than that of the pions in the TFR. Without particle identification the existing data cannot be used for comparison with our result on the $p/\pi$ ratio. Without $p_T$ determination it is not feasible to relate the $x$ distribution to the experimental $\eta'$ distribution. Nevertheless, on theoretical grounds we have shown why the production of hadrons in the TFR is not forbidden by momentum conservation.

\end{abstract}

\section{Introduction}

One of the interesting features of hadron production in heavy-ion collisions
(HIC) is the discovery of the scaling property of the pseudorapidity
distribution in the fragmentation region over an order of magnitude of
variation in the collision energy \cite{igb,phob}.  Such a property was
hypothesized in \cite{bcyy} for $pp$ collision, and was referred to as
limiting fragmentation.  The independence on energy, however, does not
imply independence on the collision system.  The pseudorapidity
distributions in the fragmentation region show definitive dependence on
the centrality of Au+Au collisions for any fixed energy \cite{phob,phob2}. 
Indeed, one cannot expect on theoretical grounds that the properties of
hadron production in that region should be the same in $pp$ and $AA$
collisions.  Whereas in $pp$ collisions no particle can be produced with a
rapidity  greater than the beam rapidity, that is not the case in $AA$
collisions.  In fact, in terms of the variable $\eta' = \eta -y_{\rm beam}$,
where $\eta$ is the pseudorapidity and $y_{\rm beam}$  the beam
rapidity, PHOBOS data indicate that the charge-particle distribution
$dN_{\rm ch}/d\eta'$ does not vanish in the $\eta' > 0$ region \cite{phob,phob3}. 
We shall refer to that region as the trans-fragmentation region (TFR).  In
this paper we describe the physics of hadron production in the TFR in the
framework of the recombination model.

Limiting fragmentation is a natural consequence of any formalism that
uses momentum fractions of partons as the essential variables to describe
hadronization since c.m. energy does not appear explicitly.  However,
particle production in the TFR is intriquing because it seems to violate
momentum conservation, as it certainly would in $pp$ collision.  In $AA$
collisions there are complications due to the fragments of the
non-interacting spectators that can get into the detectors at small angles. 
For peripheral collisions at relatively low energy ($\sqrt{s} = 19.6$ GeV)
$dN_{\rm ch}/d\eta'$ seems to approach a constant value as $\eta' \to 2$ \cite{phob,phob3}.  Such
effects seem to diminish at higher $\sqrt{s}$.  Let us put aside such issues,
since our interest here is in the hadronization of the interacting part of
nuclear collisions.  For that, there is at present no data in the TFR for
$\sqrt{s} =$ 200 GeV, say.  Nevertheless, it is an interesting and important
question to ask whether there exists theoretical reasons for hadrons to be
produced in the TFR. 

Since the problem deals with low-$p_T$ physics,  one cannot make use of
pQCD with any degree of confidence.  Nevertheless, if one takes the point
of view that hadrons are produced in the fragmentations region (FR) by
parton fragmentation, such as in the dual parton model \cite{cstt}, then no
hadrons can appear in the TFR, since all partons have momentum fractions
$x < 1$.  On the other hand, in parton recombination the momentum
fractions add and can result in a hadron momentum fraction greater than
1, provided that the constituents come from different nucleons in the
colliding system, a condition that is readily satisfied in $AA$ collisions.  It
is with that possibility in mind that we study in detail the problem of hadron production in the TFR in the
recombination model (RM) \cite{dh,rch,hy}.

Nuclear collisions have the complication of momentum degradation of
constituents traversing nuclear matter.  We have investigated the
degradation effect in $pA$ collisions, treating the constituents in terms of
valons \cite{rch,hy2}, and found good agreement with the data on ``baryon
stopping'' in the FR \cite{hy3,hgf}.  Here, in $AA$ collisions the medium is
dense, and our focus is in the TFR.  Nevertheless, similar formalism will be
used to take the degradation effect into account.

Since hadronization is a problem that involves the momenta of the
consituents in an essential way (as does momentum degradation), we shall
be working with the momentum fraction variable $x$, instead of the
pseudorapidity variable $\eta$.  In terms of $x$ the TFR is more precisely
$x > 1$, and the FR is for $x < 1$, but above, say, 0.2.  These regions do not
map isomorphically to regions in $\eta$, since $p_T$ is involved in the
definition of angle $\theta$.  Although $\eta$ is a more convenient
variable for experimental detection, we shall work entirely in the $x$
variables.  The mismatch between theory and experiment in that respect
can be overcome only when the $p_T$ values of the detected particles are
determined.  Until then, we cannot compare our predictions with any
data.  In our treatment the physics of hadronization on both sides of $x = 1$
 is continuous, so we shall calculate the spectra in the region $0.6 <x
< 1.2$ which we broadly refer to as TFR, as it roughly corresponds to the
$\eta' >0$ region (although in principle, there is no upper limit on $x$).

Apart from details our main qualitative prediction is that protons dominate
the charge particles detected in the TFR.  Such a prediction should be easier
to confirm or falsify than the verification of our results on the inclusive
distributions in $x$.  It is hoped that this paper will stimulate an effort to
improve particle idenfication in the TFR.

\section{Preliminary Considerations}

Let us begin with the kinematics $x$ relevant for the TFR in HIC at
high energy.  For hadrons detected at small angle $\theta$ relative to the
beam axis, we may approximate $\tan \theta/2$ by $p_T/2p_L$, where
$p_T$ and $p_L$ are the transverse and longitudinal momenta,
respectively, of the produced hadron.  Thus the pseudorappidity is $\eta =
\ln (2p_L/p_T)$, while the beam rapidity is $y_{\rm beam} = \ln
(\sqrt{s}/m_p)$, where $m_p$ is the proton mass.  With $\eta'$ being the
shifted pseudorapidity, $\eta' = \eta - y_{\rm beam}$, we have for
Feynman $x$, defined by $x = 2 p_L /\sqrt{s}$, 
\begin{eqnarray}
x = {p_T \over m_p} e^{\eta'} \ .
\label{1}
\end{eqnarray} 
The mapping between $\eta'$ and $x$ therefore depends on $p_T$.  If in
the forward region $\left<p_T\right><m_p$, then the $\eta' > 0$ region corresponds to
a range in $x$ that straddles $x = 1$.  For that reason we study the hadron
spectra in the range $0.6 < x < 1.2$ as a representative of the TFR.

Next we consider the geometrical aspect of nuclear collisions.  For
convenience in identifying the forward direction, let us consider $AB$ 
collision, and we are interested in the production of hadrons in the TFR of
$A$.  In the Glauber model we write for the cross section of $AB$ collision in
the form 
\begin{eqnarray}
\sigma ^{AB} = \int d^2b\, g^{AB} (b) ,
\label{2}
\end{eqnarray} 
where
\begin{eqnarray}
g^{AB} (b) &=& \int d^2s \, T_A(s) \left[1 - e^{-\sigma T_B(|\vec{s}-\vec{b}|)}
\right]\nonumber\\
&&+ \int d^2s \, T_B(s) \left[1 - e^{-\sigma T_A(|\vec{s}-\vec{b}|)}
\right] .
\label{3}
\end{eqnarray} 
$T_A(s)$  is the thickness function normalized to $A$, i.e., 
\begin{eqnarray}
T_A(s) = A \int dz\,\rho (s, z) , \qquad  \int d^2s\,T_A(s) = A,
\label{4}
\end{eqnarray}
$\rho$ being the nuclear density normalized to 1; $\sigma$ in Eq.\
(\ref{3}) is the inelastic nucleon-nucleon cross section.

To calculate the average number of wounded nucleons in $B$, we consider
only the first term in Eq.\ (\ref{3}) and call it $g^{(A)B} (b) $.  Define
$\Pi_{\nu}^{(A)B} (b)$ to be the probability of $A$ having $\nu$ collisions
in $B$ at impact parameter $b$, so that 
\begin{eqnarray}
g^{(A)B} (b) = \sum^B_{\nu = 1} \Pi^{(A)B}_{\nu} (b) \ . 
\label{5}
\end{eqnarray}
Then we have 
\begin{eqnarray}
\Pi^{(A)B}_{\nu} (b) = \int d^2s\ T_A(s) \pi^{pB}_{\nu}(|\vec{s}-\vec{b}|) \ ,
\label{6}
\end{eqnarray} 
where $\pi^{pB}_{\nu}$ is the corresponding probability in $pB$ collision
\cite{hw}
\begin{eqnarray}
\pi^{pB}_{\nu}(b) = {1 \over \nu!} \left[\sigma T_B (b) \right]^{\nu}
\exp \left[- \sigma T_B (b) \right]  \ ,
\label{7}
\end{eqnarray}
from which one can recover the necessary condition 
\begin{eqnarray}
\sum^B_{\nu=1}  \pi^{pB}_{\nu}(b)  = 1-\exp \left[- \sigma T_B (b) \right]   \ ,
\label{8}
\end{eqnarray}
as required by Eqs. (\ref{5}) and (\ref{6}).  Moreover, we obtain
\begin{eqnarray}
\sum^B_{\nu=1} \nu   \pi^{pB}_{\nu} (b)  = \sigma T_B (b)   \ .
\label{9}
\end{eqnarray}
Thus for $pB$ collision the average number of wounded nucleons in $B$ at
$b$ is
\begin{eqnarray}
\bar{\nu}^{pB}(b) = {\sum^B_{\nu = 1} \nu \pi^{pB}_\nu(b)\over \sum^B_{\nu =
1} 
\pi^{pB}_\nu(b)}= {\sigma T_B (b) \over 1 - \exp [-\sigma T_B(b)]}  \ .
\label{10}
\end{eqnarray} 

Now, returning to $AB$ collision the average number of wounded nucleons in
$B$ is
\begin{eqnarray}
\bar{\nu}^{(A)B} (b) \equiv {\sum^B_{\nu=1} \nu   \Pi^{(A)B}_{\nu}(b) \over
\sum^B_{\nu=1}   \Pi^{(A)B}_{\nu}(b)} = {\sigma \int d^2s\ T_A(s) T_B
(|\vec{s}-\vec{b}|)\over g^{(A)B}(b)} \ .
\label{11}
\end{eqnarray}
Since the number of binary collision, $N^{AB}_{\rm coll}(b)$, is the
numerator of the last expression in Eq.\ (\ref{11}), and the number of
participants, $N^{AB}_{\rm part}$, is $g^{AB}(b)$, we have for $AA$
collisions
\begin{eqnarray}
\bar{\nu}^{(A)A} (b) = {N^{AA}_{\rm coll} (b)\over N^{AA}_{\rm part}
(b)/2} \ .
\label{12}
\end{eqnarray}
For Au+Au collisions at 200 GeV, the tabulated values of $N_{\rm coll}$
and $N_{\rm part}$ are 1065 and 351 (at 0-5\% centrality), and are 220
and 114 (at 30-40\%) \cite{ssa}.  Hence we get
\begin{eqnarray}
\bar{\nu}^{(A)A}& = &6.1 \qquad  (0-5\%) \ , \nonumber\\
&&3.9 \qquad (30-40\%) \ .
\label{13}
\end{eqnarray} 

The impact parameters that correspond to the two centrality bins can be
calculated from the overlap function
\begin{eqnarray}
T_{AA} (b) = \int d^2s\,T_A(s) \,  T_A (|\vec{s}-\vec{b}|) \ .
\label{14}
\end{eqnarray}
Using the simplified form for $\rho (s,z)$ with uniform density in Eq.\
(\ref{4}) results in a distribution for $T_{AA} (b)$ that is slightly higher
than the tabulated values for various centrality bins given in \cite{ssa}.  Nevertheless, from the shape of the distribtion the corresponding values of $b$ can reasonably be set at 
\begin{eqnarray}
b &=&1\ {\rm fm} \quad  (0-5\%)\nonumber\\
&&8\ {\rm fm} \quad (30-40\%) \ .
\label{15}
\end{eqnarray}

The consideration of nuclear geometry and the associated wounded
nucleons will become important in the following when the momentum
degradation effect is to be taken into account. 

\section{Hadron Production at Large $p_L$ in the Recombination Model}

In $pp$ collisions no particle can be produced with $p_L >\sqrt{s}/2$.  In
$AB$ nuclear collisions one would initially expect the same to be true also. 
However, $n$ nucleons in $A$ have a combined momentum of $n
\sqrt{s}/2$, which can make possible a particle produced with $p_L >
\sqrt{s}/2$ without violating momentum conservation, if a coherence
effect is at work. That would be the case if the valence quarks in three
nucleons in $A$, each with momentum fraction $x_i > 1/3$, say, recombine
to form a nucleon, whose Feynman $x$ can then exceed 1.  It is also
possible for a pion to be produced with $x>1$, but since an antiquark in the
sea (or a gluon converted to $q\bar{q}$) is needed with large enough $x_i$
to recombine with a valence quark of another nucleon, the probability is
much lower.  This type of consideration need not be restricted to the $x>1$
region.  Even for $x>0.6$, which we broadly refer to as TFR, the
recombination of partons from different nucleons in $A$ will dominate
over those processes where the partons are from the same nucleon, as in
$pB$ collisions.  That dominance is over and above the shifted peak in
rapidity due to ``baryon stopping'' simply on the basis of extra momentum
availability in $AB$ collisions.  Thus even without detail calculations we can
predict that $p/\pi$ ratio is large in the TFR in $AB$ collision.  Similarly, $\Lambda/K$
ratio is also large.

Hadron production at low $p_T$ and large $p_L$ in  hadronic collisions has been treated in
the RM in good agreement with data \cite{dh,rch,hy}.  The extension now
to nuclear ($AB$) collisions with emphasis in the TFR region has the same
basic recombination formula 
\begin{eqnarray}
H^{AB}_p (x) \equiv x {dN^{AB}_p\over dx}
= \int {dx_1\over x_1} {dx_2\over x_2} {dx_3\over x_3}F^{AB}_{uud}(x_1, x_2, x_3)R_p
(x_1, x_2, x_3, x)
\label{16}
\end{eqnarray}
for the production of proton.  $R_p$ is the recombination function (RF) that
has been studied in the framework of the valon model for the nucleon
structure \cite{rch,hy2}, by relating $R_p$  to the valon distribution
$G_p$ in the proton \cite{hy4}  
\begin{eqnarray}
R_p (x_1, x_2, x_3, x) = g_{st}{x_1 x_2 x_3\over x^3}  G_p \left({x_1 \over
x}, {x_2 \over x},  {x_3 \over x}\right) \ ,
\label{17}
\end{eqnarray}
where 
\begin{eqnarray}
G_p (y_1, y_2, y_3) = g_p(y_1 y_2)^{\alpha}y^{\beta}_3 \delta (y_1 +
y_2 + y_3-1) ,
\label{18}
\end{eqnarray} 
\begin{eqnarray}
g_p = \left[B (\alpha +1, \alpha +
\beta +2) B (\alpha +1, \beta +1)
\right]^{-1},  
\label{19}
\end{eqnarray}
and $g_{st}$ is the statistical factor $1/6$ \cite{hy4}.  It is by successfully
fitting the CTEQ parton distribution functions at low $Q^2$ that the
parameters $\alpha$ and $\beta$ are determined to be \cite{hy2}
\begin{eqnarray}
\alpha = 1.75, \qquad \beta = 1.05 .
\label{20}
\end{eqnarray}
The $\delta$  function in Eq.\ (\ref{18}) enforces the momentum sum for
recombination:  $\sum_ix_i = x$.

The key quantity in Eq.\ (\ref{16}) is the 3-quark distribution
$F^{AB}_{uud} (x_1, x_2, x_3)$.  If it were like in $pB$ collision, then the
quarks would all originate from the projectile, and the $x_i$ in
$F^{pB}_{uud} (x_1, x_2, x_3)$ would satisfy $\sum_i x_i < 1$.  However,
in the TFR of $AB$ collisions we consider the dominant component where
each quark is from a separator nucleon in the same longitudinal tube at
distance $s$ from the center of $A$, so that we can write in the factorizable
form
\begin{eqnarray}
F^{(3)B}_{uud} (x_1, x_2, x_3) = F^u_{\bar{\nu}}(x_1)
F^u_{\bar{\nu}}(x_2)F^d_{\bar{\nu}}(x_3) 
\label{21}
\end{eqnarray}
with the labels $\vec{s}$ and $\vec{b}$ suppressed.   On the LHS of Eq.\
(\ref{21}) we use (\ref{3}) instead of $A$ in the superscript to emphasize
that only 3 nucleons in $A$ are considered.  How to generalize from 3 to
$A$ will be discussed in Sec. IV.  The effect of momentum degradation due
to the passage through $B$ will be considered in the next section. The
crucial point here is that the variables $x_i$ in Eq.\ (\ref{21}) are
independent of one another, so the integrals in Eq.\ (\ref{16}) are from 0
to 1 for each $x_i$.  Thus the maximum possible $x$  is 3, well beyond  the
conventional FR.  It is this unconventional possibility of producing a proton
in the TFR that motivates our investigation here.

In Eq.\ (\ref{21}) we have used the superscripts $u$ and $d$ to denote the
flavors of quarks that are to recombine in Eq.\ (\ref{16}) where the RF is
given by Eqs.\ (\ref{17}) and (\ref{18}) with $y_1$ and $y_2$ referring to
the $u$ quark and $y_3$ to the $d$ quark.  However, if the projectile $A$
is an isoscalar which we shall assume, then at every impact parameter $s$
there are equal numbers of protons and neutrons, so $F^u_{\bar{\nu}} =
F^d_{\bar{\nu}}$.  In the valon model for $pB$ collisions the quark
distributions are \cite{hy3}
\begin{eqnarray}
F^{u,d}_{\bar{\nu}}(x_i) = \int^1_{x_i} dy' \bar{G}'_{\bar{\nu}} (y') K
\left({x_i \over y'}\right) \ ,
\label{22}
\end{eqnarray}
where the valon distribution $\bar{G}'_{\bar{\nu}} (y')$ differs from
$G(y)$ due to momentum degradation, as will be discussed in detail in the
following section.  $K(z)$ is the quark distribution in a valon.  Since our 
valon distribution is flavor independent, $K(z)$ consists of both valence
and sea quarks \cite{hy2}
\begin{eqnarray}
K(z) = K_{NS}(z) + L(z)  \ ,
\label{23}
\end{eqnarray}
where
\begin{eqnarray}
K_{NS}(z) = z^a (1-z)^b/B(a,b+1)  \ ,  \quad a = 0.35 , \quad b = -0.61 \ .
\label{24}
\end{eqnarray}
These parameters $a$ and $b$ are determined from the moments of
$K_{NS}$ given in \cite{hy2}.  For the sea quark distribution $L(z)$
we have to do two things:  first, we have to average the favored $(L_f)$
and unfavored $(L_u)$ distributions given \cite{hy2}; second, we readjust
the normalization to saturate the sea.  $L_f(z)$ and $L_u(z)$ have been
determined for the proton sea to distinguish, for example, a $u$ quark in
a $U$ valon (favored) from a $d$ quark in a $U$ valon (or a $u$ quark in
a $D$ valon, unfavored).  They are, for $Q = 1$ GeV/c, \cite{hy2}
\begin{eqnarray}
\ln L_f(z) &=&   -2.66 + 0.08t - 10.4 t^2 -6t^3\nonumber\\
 \ln L_u(z) &=&   -2.92 + 4.0t - 5.95 t^2 -1.4t^3 \ ,
\label{24a}
\end{eqnarray}
where $t = -\ln (1-z)$.  We define
\begin{eqnarray}
L_q(z) = {1 \over 2}\left[L_f(z)+ L_u(z)\right] \ .
\label{25}
\end{eqnarray}
This distribution does not include the conversion of gluons to the sea
quarks, a process that we must consider in order to account for the
hadronization of all partons, including gluons \cite{rch,hy}.  Thus to
saturate the sea, we renormalize as follows
\begin{eqnarray}
L'_q (z) =  Z L_q (z)   \ ,
\label{26}
\end{eqnarray}
where $Z$ is determined by solving the two algebraic equations that
express the momentum conservation in the $u, d, s$ sections in terms of
the second moments \cite{hy}
\begin{eqnarray}
\tilde{K}_{NS}(2) + 2 \left[2\tilde{L}_q(2) + \tilde{L}_s(2) 
\right] +\tilde{L}_g(2) =1
\label{27}
\end{eqnarray}
\begin{eqnarray}
\tilde{K}_{NS}(2) + 2 \left[2\tilde{L}'_q(2) + \tilde{L}'_s(2) 
\right]=1
\label{28}
\end{eqnarray}
\begin{eqnarray}
Z  = 1 +{\tilde{L}_g(2) \over 2 \left[ 2 \tilde{L}_q(2) +  \tilde{L}_s(2)\right]
} \ .
\label{29}
\end{eqnarray}
In Eq.\ (\ref{28}) we have assumed that the $s$-quark sea is enhanced
also in nuclear collisions, unlike the case of hadronic collision where the
gluons are convected to the light-quark sector only through $g \to q
\bar{q}$  \cite{hy}.  In getting Eq.\ (\ref{29}) we have set $L'_s = Z L_s$ 
also.  Since the second moments in Eq.\ (\ref{29}) are tabulated in 
\cite{hy}, we obtain 
\begin{eqnarray}
Z = 3.42 \ .
\label{30}
\end{eqnarray}
In the following we shall consider $L'_q (z)$ only whenever the light
quarks in the sea are needed, including the $L(z)$ term in Eq.\
(\ref{23}).

For pion production the physical content of the calculational procedure is
the same as for proton production, except that it is a $q\bar{q}$
recombination.  Thus, as in Eq.\ (\ref{16}), (\ref{21}) and (\ref{22}) we
have
\begin{eqnarray}
H^{AB}_{\pi}(x)\equiv x {dN^{AB}_{\pi}  \over  dx} = \int {dx_1  \over 
x_1}{dx_2 \over  x_2} F^{AB}_{q\bar{q}} (x_1, x_2) R_{\pi}(x_1, x_2, x) \ , 
\label{31}
\end{eqnarray}
\begin{eqnarray}
F^{(2)B}_{q\bar{q}} (x_1, x_2) = F^q_{\bar{\nu}} (x_1)  F^{\bar q}_{\bar{\nu}} (x_2)
 \ ,
\label{32}
\end{eqnarray}
\begin{eqnarray}
F^{\bar q}_{\bar{\nu}} (x_2) = \int^1_{x_2} dy' \bar{G}_{\bar{\nu}} (y')L'_q
\left({x_2  \over y'}\right) \ ,
\label{33}
\end{eqnarray}
where $F^{\bar{q}}_{\bar{\nu}}$ is obtained from the saturated sea.  The
recombination function for pion is \cite{rch,hy}
\begin{eqnarray}
R_{\pi}(x_1, x_2, x) = {x_1 x_2  \over x^2}\,\delta\left({x_1 \over x}
+ {x_2  \over x} -1 \right)
\label{34}
\end{eqnarray}
In an isosymmetric collision system we need not distinguish the charge
states.  $\Lambda$ and $K$ production can similarly be considered.

\section{Momentum Degradation}

In the preceding section we have described the quark distribution in Eq.\
(\ref{22}) as a convolution of the valon distribution and the quark
distribution in a valon, but we have not specified the former.  In a $pp$
collision the valon distribution $G(y)$, $y$ beginning the momentum
fraction, has been identified as that of the free proton on the basis that at
low $p_T$ and large $p_L$ the fast partons in the forward direction are
unaffected by the opposite-going partons due to the lack of long-range
correlation in rapidity.  The valon model connects the bound-state
problem of a static proton (in terms of constituent quarks) with the
structure problem of a proton in collision (in terms of partons) \cite{rch}. 
In $pA$ collision the effect of momentum degradation in the passage of
the projectile through the nuclear target is applied to the valons in Ref.\
\cite{hy3}, where it is shown not only how baryon stopping can be
obtained in agreement with data, but also how pion production in the
FR can be determined.  We now extend the treatment to the TFR in the
$AB$ collisions.

In a free proton the single-valon $U$ and $D$ distributions are obtained
from the exclusive distribution given in Eq.\ (\ref{18}) by integration
\begin{eqnarray}
G^U(y_1) &=& \int^{1-y{_1}}_0 dy_2 \int^{1-y_1-y_2}_0 dy_3\,G_p (y_1, y_2, y_3)
\nonumber\\
&=& g_p B(\alpha + 1, \beta +1) y^{\alpha}_1(1 - y_1)^{\alpha + \beta +
1}
\label{35}
\end{eqnarray}
\begin{eqnarray}
G^D(y_3) &=& \int^{1-y{_3}}_0 dy_1 \int^{1-y_1-y_3}_0 dy_2\,G_p (y_1,
y_2, y_3)
\nonumber\\
&=& g_p B(\alpha + 1, \alpha +1) y^{\beta}_3 (1 - y_3)^{2 \alpha  +
1} \ .
\label{36}
\end{eqnarray}
For isoscalar nuclei we take the average
\begin{eqnarray}
G(y) = {1  \over  2} \left[ G^U (y) + G^D (y)\right]
\label{37}
\end{eqnarray}
before the nucleon traverses the nuclear medium.  We note that
$G(y)$ is not an invariant distribution, but $yG(y)$ is.  It is
normalized by
\begin{eqnarray}
\int^1_0 dy G(y) = \int dy_1dy_2dy_3\,G(y_1,y_2,y_3)=1 \ ,
\label{38}
\end{eqnarray}
i.e., the probability for the proton to consist of three and only three
valons is 1.

Suppose now that a nucleon in $A$ at a fixed impact parameter $s$
collides with $\bar{\nu}$ wounded nucleons in $B$ on the average. 
Since $|\vec{s}- \vec{b}|$ can be almost as large as the radius $R$, it
is possible for $\bar{\nu}(b,s)$ to be very small.  That is the
geometrical situation where several nucleons in $A$ can contribute 
to the TFR without much momentum degradation.  When we consider
fluctuations of $\nu$ from $\bar{\nu}$, we must include the
possibility of $\nu$ being $0$.  Thus we use the Poisson distribution
\begin{eqnarray}
P_{\bar{\nu}}(\nu) = {\bar{\nu}^{\nu}  \over  \nu!} e^{-\bar{\nu}}
\label{39}
\end{eqnarray}
with normalization defined by summation from $\nu = 0$  
\begin{eqnarray}
\sum^{\infty}_{\nu = 0} P_{\bar{\nu}} (\nu) = 1 \ .
\label{40}
\end{eqnarray}

If a valon loses a momentum fraction $1 - \kappa$ at each collision,
then after $\nu$ collisions the modified valon distribution is
\begin{eqnarray}
y'G'_{\nu} (y') = \int^1_{y'} dy\,G(y)\,\delta \left({y'\over y} - \kappa^{\nu} \right)     \ ,
\label{41}
\end{eqnarray}
from which follows
\begin{eqnarray}
G'_{\nu} (y') = \kappa^{-2\nu} G(\kappa ^{-\nu}y') \ .
\label{42}
\end{eqnarray}
The parameter $\kappa$ for the reduced momentum fraction is
unknown, since it should not be inferred from $pA$ collision, which is
for a cold nuclear target.  In $AB$ collision a tube in $A$ contains
many nucleons which cannot all be treated as if each of them
collides with $\nu$ nucleons in a cold nucleus $B$.  Indeed, it is hard
to assess the state of $B$ when the back part of the tube traverses
the medium.  We can at best use an adjustable parameter  $\kappa$
to describe in some average sense what contributes to the TFR. 
From Eq.\ (\ref{42}) we get after $\bar{\nu}$ collisions on the
average
\begin{eqnarray}
\bar{G}'_{\bar{\nu}} (y') = \sum^{\infty}_{\nu = 0} G'_{\nu} (y')
P_{\bar{\nu}}(\nu) \ .
\label{43}
\end{eqnarray} 
To identify $\bar{\nu}$ with the quantity expressed by Eq.\
(\ref{10}) is based on the assumption that each valon experiences the
same average number of collisions as the parent nucleon does.

$\bar{G}'_{\bar{\nu}} (y')$, as given in Eq.\ (\ref{43}) is the modified
valon distribution that should be used in Eqs.\ (\ref{22}) 
and (\ref{33}).  That takes care of the nuclear effect.  What remains
is the calculation of the $p$ and $\pi$ distributions in the TFR.

\section{Proton and Pion Distributions}

To calculate the hadron distributions in $x$, we return to Eq.\
(\ref{16}) as a general formula, which needs, however, some more
elaboration to account for the size of $A$.  In Eq.\ (\ref{21}) we show
the factorizable form of $F^{(3)B} _{uud}$ when there are only
three nucleons in the projectile, each contributing a quark.  Now, we
specify the details of how to calculate $F^{AB} _{uud}$  that is
called for in Eq.\ (\ref{16}).

In Sec.\ II we have in Eq.\ (\ref{7}) the probability
$\pi^{pB}_{\nu}(b)$ for a nucleon making $\nu$ collisions in $B$ at
impact parameter $b$.  Now, consider the same quantity in $A$ and
write
\begin{eqnarray}
\pi^{Ap}_{\mu} (s) = {1  \over  \mu!} \left[\sigma T_A (s)
\right]^{\mu} \exp \left[- \sigma T_A (s) \right]
\label{44}
\end{eqnarray}
for the probability that $\mu$ nucleons in $A$ colliding with a
nucleon in $B$ at impact parameter $s$ in $A$.  In place of 
Eq.\ (\ref{6}) we now have for $\mu$ nucleons in $A$ colliding with
$\nu$ nucleons in $B$ the probability
\begin{eqnarray}
\Pi^{AB}_{\mu \nu}(b) = \int {d^2s  \over  \sigma} \pi^{Ap}_{\mu}(s) \pi^{pB}_{\nu}
(|\vec{s}- \vec{b}|) \ .
\label{45}
\end{eqnarray}
Since at least 3 nucleons  in $A$ are needed for our calculation of the proton distribution in the TFR of $A$, as done in
Sec. III, we must sum over $\mu$ starting with $\mu = 3$ and get
\begin{eqnarray}
H^{AB}_p (x,b) = \int {d^2s  \over  \sigma}
\sum^{\infty}_{\mu = 3}\pi^{Ap}_{\mu} (s) {\mu \choose 3}
H^{(3)B}_p (x, b, s) \ ,
\label{46}
\end{eqnarray}
where $H^{(3)B}_p (x, b, s)$ is what we have described in Eqs.\ (\ref{16}), (\ref{21}) and (\ref{22}), i.e., 
\begin{eqnarray}
H^{(3)B}_p (x, b, s) = \int \left[ \prod^3_{i = 1} {dx_i \over x_i}F^q_{\bar{\nu}}(x_i) \right] R_p (x_1, x_2, x_3, x)  \ ,
\label{47}
\end{eqnarray}
where $\bar{\nu}$ is given by Eq.\ (\ref{10}) but at $|\vec{s}-\vec{b}|$
\begin{eqnarray}
\bar{\nu} = \bar{\nu}^{pB}(|\vec{s}-\vec{b}|) =  {\sigma T_B (|\vec{s}-\vec{b}|) \over 1 - \exp \left[ - \sigma T_B (|\vec{s}-\vec{b}|)\right]} \ .
\label{48}
\end{eqnarray}
The sum over $\mu$ in Eq.\ (\ref{46}) can be performed, yielding 
\begin{eqnarray}
H^{AB}_p (x,b) = \int {d^2s  \over  \sigma}
{\left[\sigma T_A(s)  \right]^3\over 3!}\,
H^{(3)B}_p (x, b, s) \ .
\label{49}
\end{eqnarray}
Note that with this formula we do not need the results on $\bar{\nu}^{(A)A} (b)$ given in  Eqs.\ (\ref{12}) and (\ref{13}).

For pion production it is straightforward to modify Eq.\ (\ref{49}) and get 
\begin{eqnarray}
H^{AB}_{\pi} (x,b) = \int {d^2s  \over  \sigma}
{\left[\sigma T_A(s)  \right]^2\over 2!}  H^{(2)B}_{\pi} (x, b, s) \ ,
\label{50}
\end{eqnarray}
where
\begin{eqnarray}
H^{(2)B}_{\pi} (x,b.s) = \int {dx_1  \over  x_1} {dx_2  \over  x_2}
F^q_{\bar{\nu}}(x_1) F^{\bar{q}}_{\bar{\nu}}(x_2) R_{\pi} (x_1, x_2, x) \ ,
\label{51}
\end{eqnarray}
Since $F^{\bar{q}}_{\nu} (x_2)$ is severely damped at large $x_2$, $H^{AB}_{\pi} (x,b)$ is expected to be much more suppressed compared to $H^{AB}_p (x,b)$ in the TFR.  Nevertheless, $H^{AB}_{\pi} (x,b)$ need not vanish for $x > 1$, unlike $H^{pB}_{\pi} (x,b)$.

On the basis of Eqs.\ (\ref{49}) and (\ref{50}) we have calculated the proton and pion distributions for Au+Au collisions at $b = 1$ and 8 fm, corresponding to 0-5\% and 30-40\% centralities, according to Eq.\ (\ref{15}).  We have used the approximation of uniform nuclear density with $R = 1.2 A^{1/3}$ fm, and $\sigma = 41$ mb.  For the parameter $\kappa$ that represents the surviving valon momentum fraction after each collision, we have chosen two representative values, $\kappa = 0.8$ and $0.6$, where $\kappa = 1$ implies no momentum degradation.  The results for the invariant distribution $xd N_p/dx$ for the proton, which is just $H^{\rm {AuAu}}_p (x, b)$, are shown in Fig.\ 1 for (a) $b = 1$  fm and (b) $b = 8$  fm.  The solid (dashed) line corresponds to $\kappa = 0.8 \,  (0.6)$.  All four lines are nearly straight, i.e., exponential in $x$, smoothly throughout the TFR.  There can be other contributions to the proton production in that region due to the recombination of quarks originating from one or two nucleons, but they are so small that we ignore them.

Figure 1 shows that the $x$ distributions are suppressed when there is more momentum degradation (smaller $\kappa$), as is expected.  In the case of $pA$ collisions such a suppression would correspond to the qualitative notion of baryon stopping.  But in $AA$ collisions, instead of stopping, we have protons produced at $x > 1$.  Nevertheless, the overall normalization is lowered when there is more momentum degradation.  Thus there are two features about the inclusive distributions of the proton:  it extends smoothly beyond $x = 1$, and is more suppressed at lower $\kappa$.  The physical value of $\kappa$ that corresponds to reality can be determined only after data become available and put on plots such as Fig.\ 1.  For the purpose of summarizing the behavior of the $x$ distributions, we parametrize them in the form 
 \begin{eqnarray}
x {d N_p\over dx} (b) = \exp \left[h_0^p (b) - h_1^p (b)x  \right], \qquad 0.6 < x < 1.2  \ ,
\label{52}
\end{eqnarray}
which fits the lines in Fig.\ 1 extremely well with the parameters $h_0^p (b)$ and $h_1^p (b)$ given in Table 1.

\begin{table}[htdp]
\caption{Parameters $h_0^p$ and $h_1^p$ for proton}
\begin{center}
\begin{tabular}{|c|c|c|c|c|}\hline
&\multicolumn{2}{c|}{$\kappa = 0.8$}&\multicolumn{2}{c|} {$\kappa = 0.6$}\\
\hline
&$b = 1$&$b=8$&$b = 1$&$b=8$\\ 
\hline
$h_0^p (b)$&10.58&9.90&7.03&7.93\\
\hline
 $h_1^p (b)$&9.80&8.18&8.41&8.04\\
 \hline
\end{tabular}
\end{center}
\label{Table 1}
\end{table}

It is of interest to know how the distribution depends on centrality at a given $\kappa$.  We plot the distributions normalized by $N_{\rm{part}}/2$ in Fig.\ 2, which shows how they are suppressed as the collision changes from peripheral to central.  At larger $b$ the collisions on average have higher $\mu$ and lower $\nu$ in Eq.\ (\ref{45}), leading to more partons hadronizing in the TFR.

In \cite{phob} there are data from Au+Au collisions at 200 GeV showing that the $\eta'$ distributions for 0-6\% and 35-40\% cross over at $\eta' \approx -2$.  According to  Eq.\ (\ref{1}) that would correspond to $x \approx 0.135 p_T/m_p \stackrel{<}{\sim} 0.1$, which is significantly outside the TFR.  For $-2 < \eta' < 0$ the data show the peripheral case higher than the central case, not unlike what we have in Fig.\ 2(a).

For pion production Fig.\ 3 shows the rapid decline of the $x$ distributions, as $x$ is increased toward 1 and beyond.  Since $\bar{q}$ distribution is suppressed at large $x_2$, it is difficult for a pion to acquire enough momentum to go deep into the TFR.  Nevertheless, the $x > 1$ region is not forbidden.  It is evident from Fig.\ 3 that there is no sensitive dependence on $\kappa$.  The reason is partly because there are only two $\bar{G}'_{\bar{\nu}}$ functions in Eqs.\ (\ref{32}) and (\ref{33}) for $\pi$, instead of three for $p$, but mostly because the RF for $\pi$ is broader than that for $p$.  The valon distribution in $\pi$ is flat \cite{hy} (corresponding to pion being a tightly bound state), so the wider momentum spread allows the more degraded parton momenta to contribute to the formation of $\pi$.  In Fig.\ 4 we show the dependence on $b$.  Compared to Fig.\ 2, the pions do not show as much dependence as do the protons, although the vertical scales of the two figures are different and can lead to visual misreading.  Quantitatively, we can fit the distributions by
\begin{eqnarray}
x {d N_{\pi}\over dx} (b) = \exp \left[h_0^{\pi} (b) + h_1^{\pi} (b)x - h_2^{\pi} (b)x^2 \right], \qquad 0.6 < x < 1.2  \ ,
\label{53}
\end{eqnarray}
with the parameters given in Table II.  Qualitatively, the pion distributions are orders of magnitude lower than the proton distributions.
\begin{table}[htdp]
\caption{Parameters $h_0^{\pi}$, $h_1^{\pi}$ and $h_2^{\pi}$ for pion}
\begin{center}
\begin{tabular}{|c|c|c|c|c|}\hline
&\multicolumn{2}{c|}{$\kappa = 0.8$}&\multicolumn{2}{c|} {$\kappa = 0.6$}\\
\hline
&$b = 1$&$b=8$&$b = 1$&$b=8$\\ 
\hline
$h_0^{\pi}$&1.96&1.25&0.0063&-0.083\\
\hline
 $h_1^{\pi}$&5.46&7.39&7.40&8.32\\
 \hline
  $h_2^{\pi}$&14.15&14.79&14.66&14.87\\
 \hline
\end{tabular}
\end{center}
\label{Table II}
\end{table}

Apart from the details of the $x$ distributions, our main prediction is that proton production dominates over pion production in the TFR.  To give a visual impact of that dominance, we show in Figs.\ 5 and 6 the $p/\pi$ ratio.  At $x \sim 1$, the ratio is roughly $10^3$ for any combination of $b$ and $\kappa$.  It is such a large ratio that particle identification in the TFR would be the most direct way to settle the question whether our hadronization scheme is in any way close to reality. 

\section{Conclusion}

We have investigated hadron production in the trans-fragmentation region.  The overwhelming feature of our result is that the $p/\pi$ ratio is extremely large, roughly $10^3$ at  $x \sim 1$.  That feature is a direct consequence of parton recombination.  For a proton to be produced at  $x \sim 1$, it is rather easy to find three nucleons in $A$ each contributing a quark at $x_i \sim 1/3$ to form the proton.  However, for a pion at  $x \sim 1$, it is hard to find any antiquark at, for example, $x_i \sim 1/3$ to help a quark at $\sim 2/3$ to make up the pion momentum.  Quantitative value of the $p/\pi$ ratio can be obtained only after some rather involved calculations.  But qualitatively to have a distribution in $x$ that crosses the boundary at $x = 1$ smoothly is possible only by parton recombination, since fragmentation would require all hadrons to be produced at $x < 1$.

Experimental data do show that particles can be produced at $\eta ' > 0$, at least at lower energies.  At 200 GeV the data have stopped at $\eta ' \sim 0$, but show no evidence of vanishing there \cite{phob}.  It is unfortunate that we cannot compare our result in $x$ to the data in $\eta '$, since $p_T$ of the produced particles are unknown.  Either particle identification or $p_T$ determination, preferably both, would greatly help to relate theory and experiment.

The importance of clarifying what happens in the TFR is in the determination of whether there exists new physics in the FR and beyond.   In our treatment of the problem we have considered only low-$p_T$ physics, but extended to include recombination of quarks at medium $x_i$ from different nucleons.  If proven correct by data, it forms the basis from which to extend further to higher $p_T$ in the TFR.  Then there should arise a competition between the enhancement effect studied here and the suppression effect found earlier in forward production at intermediate $p_T$ in d+Au collisions \cite{hyf}.  The suppression can be due to either initial-state or final-state physics.  Any future study of hadron production in a larger domain in $p_T$ in the TFR will have to be consistent with the physics explored here at low $p_T$.

\section*{Acknowledgment}

We are grateful to G.\ Roland for his generous assistance in providing us with information about the PHOBOS data and to J.\ Pisut for helpful communication.  This work was supported  in
part,  by the U.\ S.\ Department of Energy under Grant No. DE-FG02-96ER40972 and by the Ministry of Education of China under Grant No. 03113.

\vskip2cm
\begin{center}
\section*{Figure Captions}
\end{center}

\begin{description}
\item
Fig.\ 1. Proton distributions in the TFR for (a) $b=1$ fm, and (b) $b=8$ fm.

\item
Fig.\ 2. Same as in Fig.\ 1 but for  (a) $\kappa=0.8$, and (b) $\kappa=0.6$, where $\kappa$ is the survival factor in momentum degradation.
 
 \item
Fig.\ 3. Pion distributions in the TFR for (a) $b=1$ fm, and (b) $b=8$ fm.
 
 \item
Fig.\ 4. Same as in Fig.\ 3 but for  (a) $\kappa=0.8$, and (b) $\kappa=0.6$. 

\item
Fig.\ 5. $p/\pi$ ratio for (a) $b=1$ fm, and (b) $b=8$ fm.

 \item
Fig.\ 6.  Same as in Fig.\ 5 but for  (a) $\kappa=0.8$, and (b) $\kappa=0.6$.

\end{description}

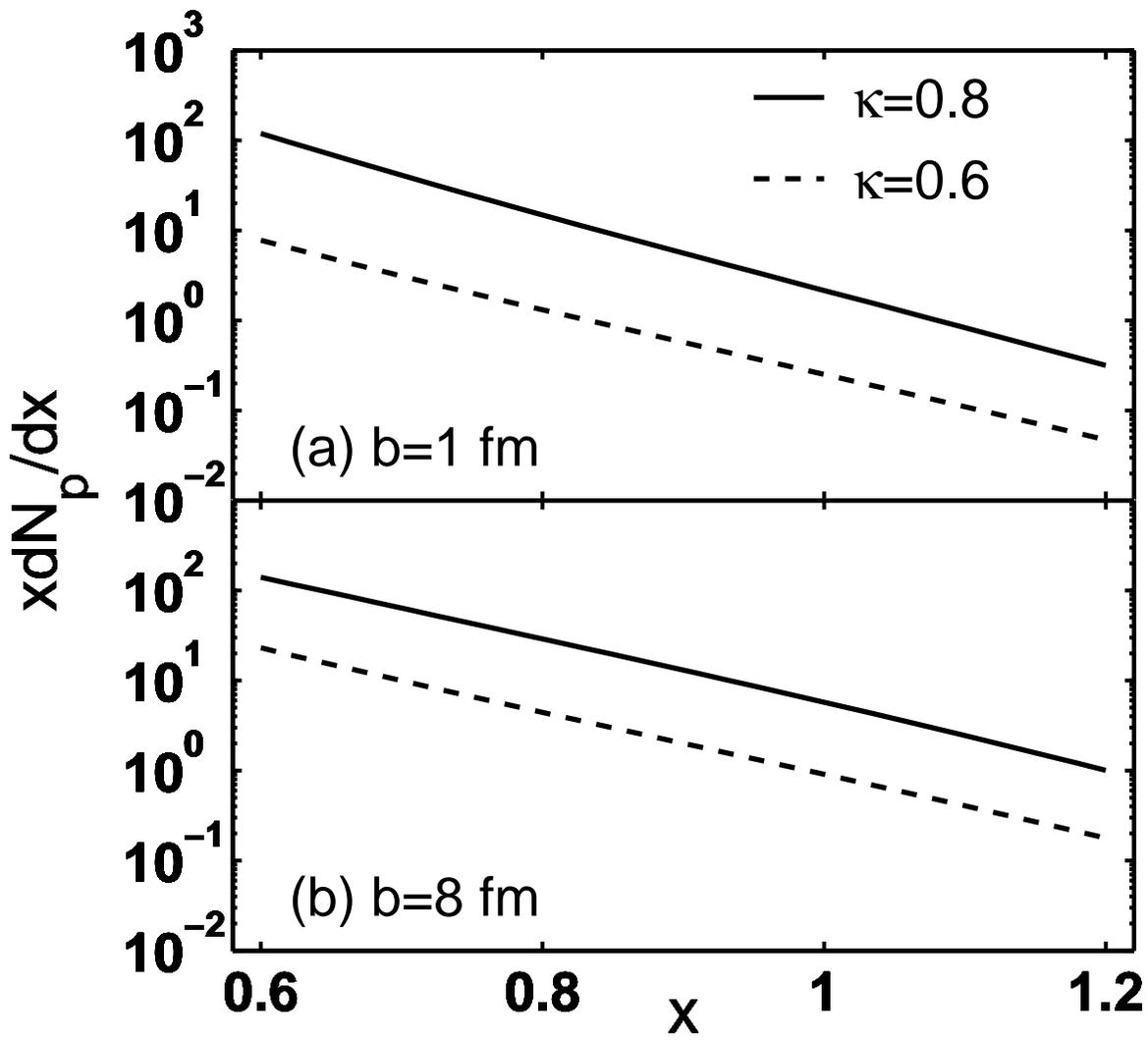
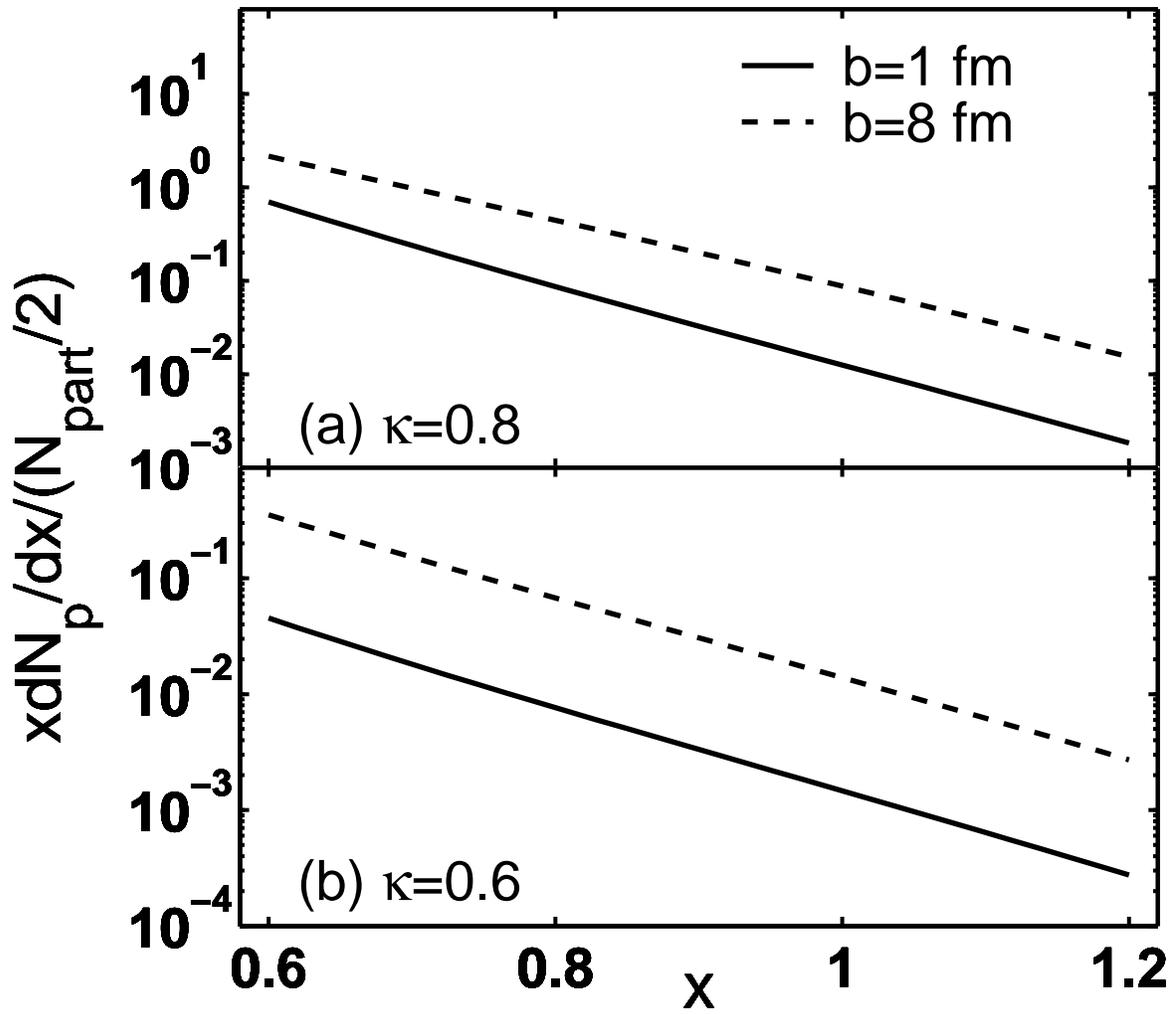
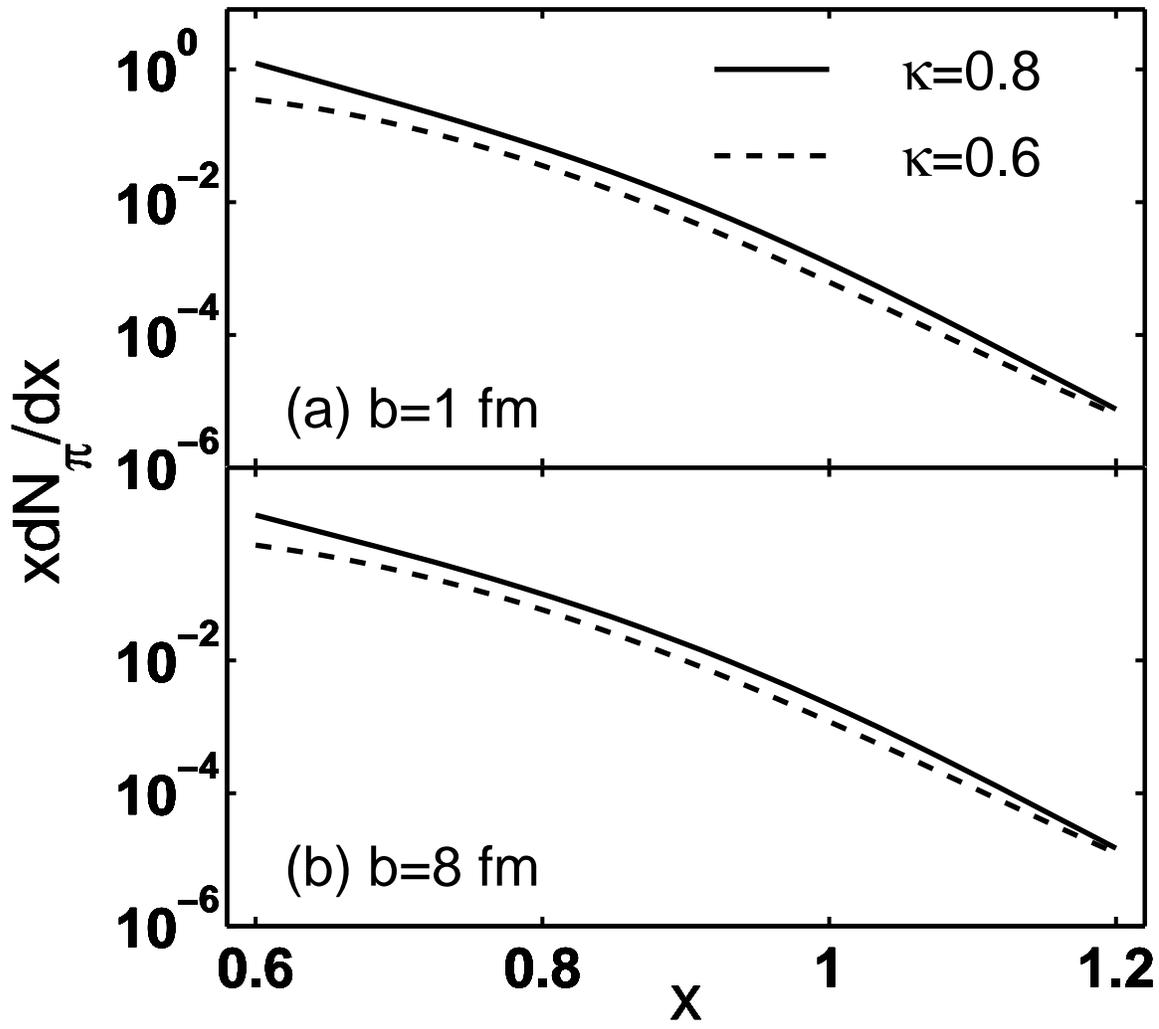
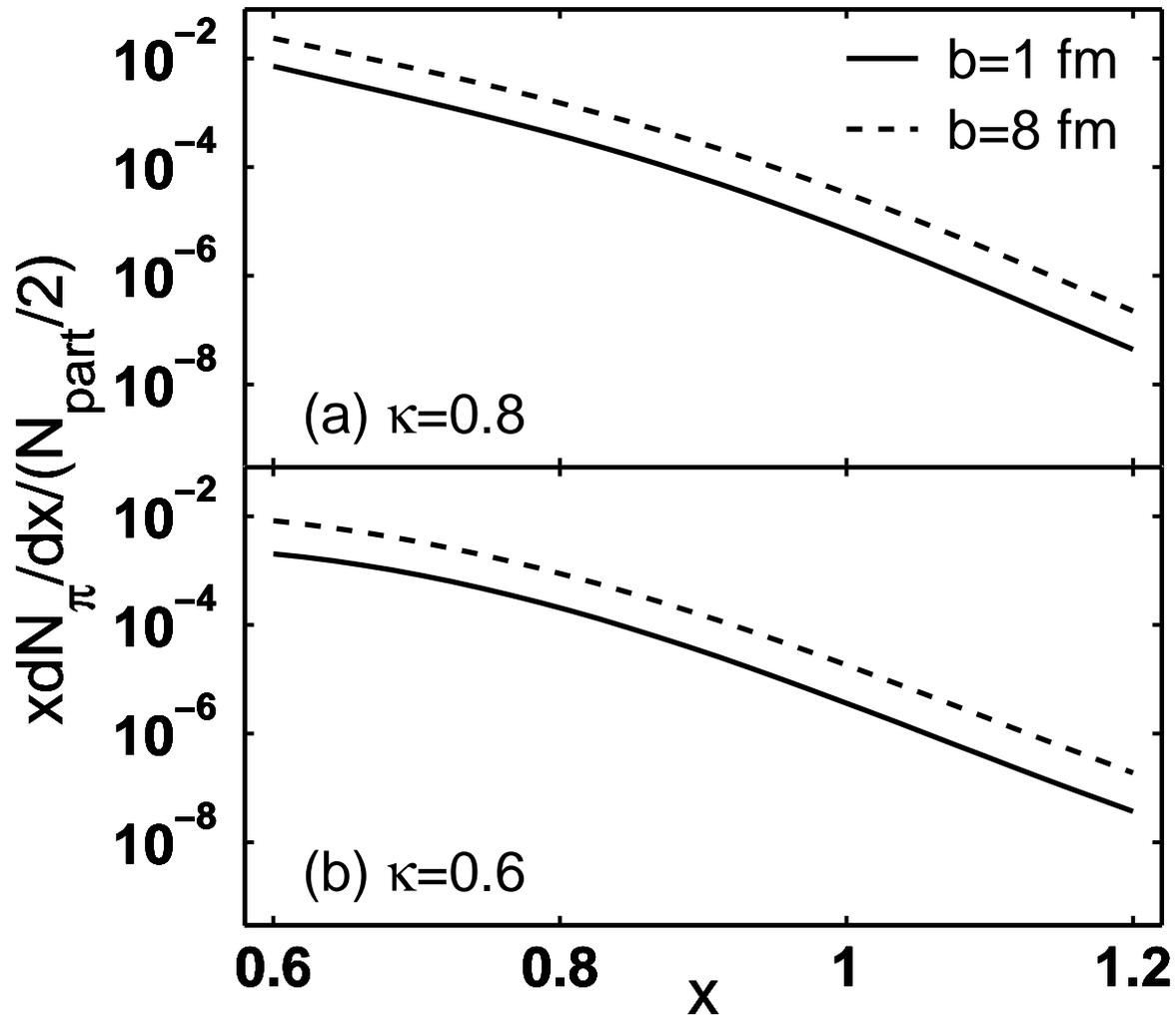
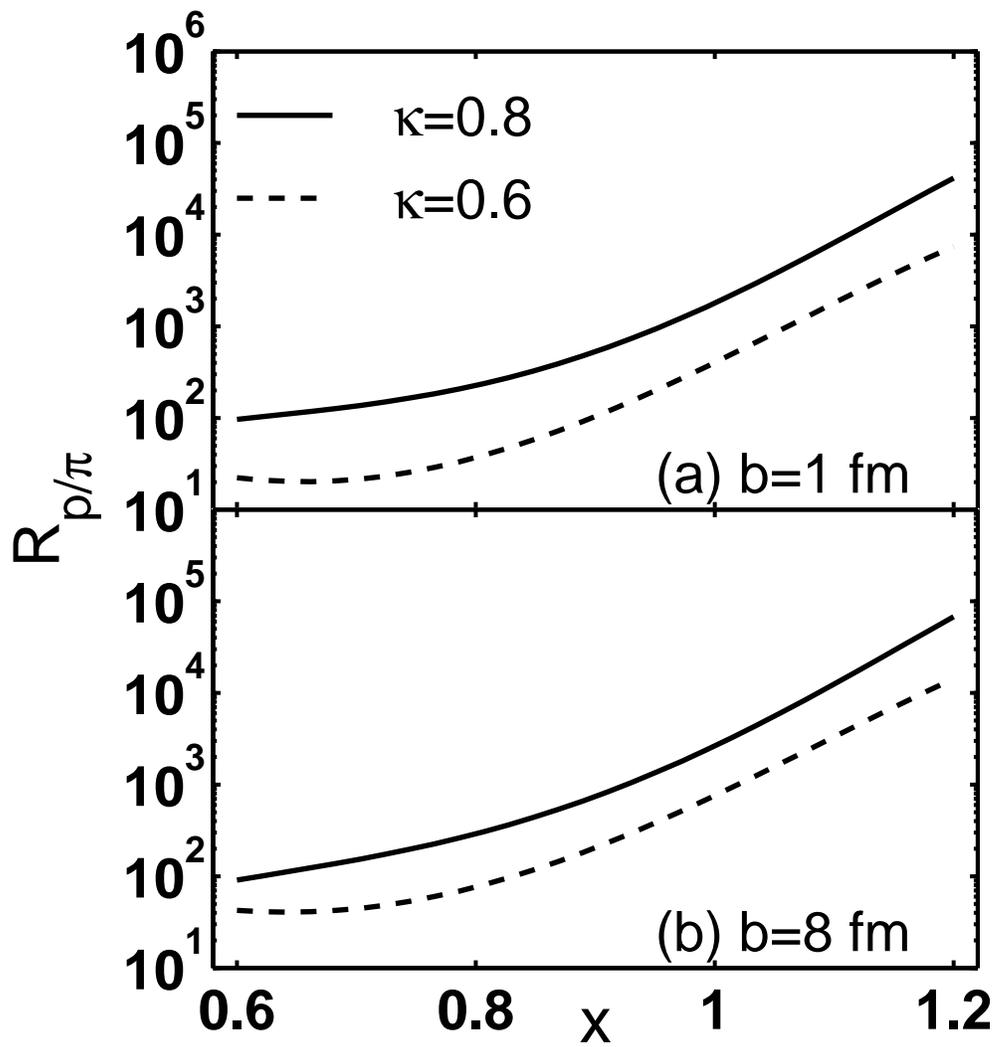
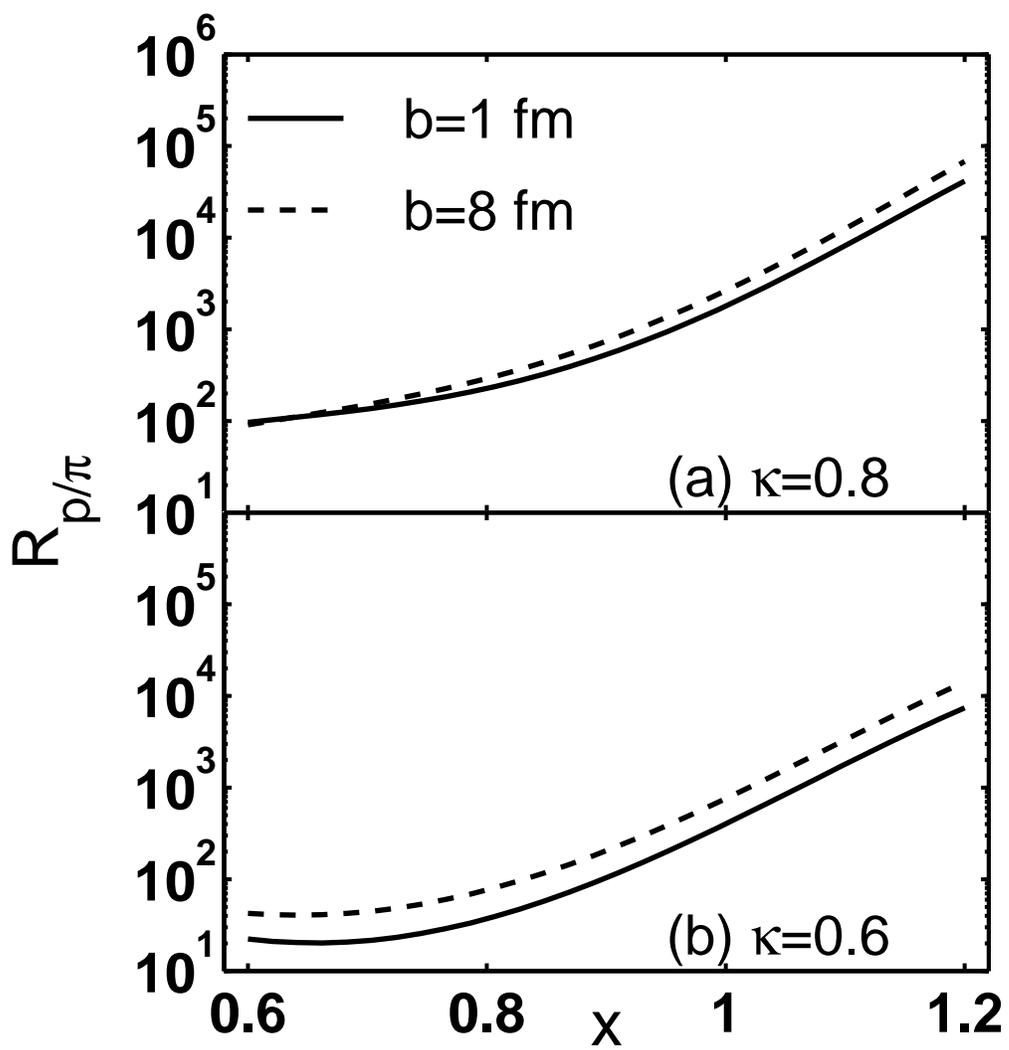
\end{document}